\documentclass[journal]{IEEEtran}

\usepackage{url}
\usepackage{graphicx}
\usepackage{amsmath}
\usepackage{amsfonts}
\usepackage{amssymb, color}

\newtheorem{theorem}{Theorem}

\begin{document}

\title{Designing Wireless Powered Networks Assisted by Intelligent Reflecting Surfaces with Mechanical Tilt}  

\author{Zoran Hadzi-Velkov, Slavche Pejoski, Nikola Zlatanov, and Haris Gačanin     
\vspace{-3mm}

\thanks{Z. Hadzi-Velkov and S. Pejoski are with the Faculty of Electrical Engineering and Information Technologies, Ss. Cyril and Methodius University, 1000 Skopje, Macedonia (email: \{zoranhv, slavchep\}@feit.ukim.edu.mk).} \thanks{N. Zlatanov is with the Department of Electrical and Computer Systems Engineering, Monash University, Clayton, VIC 3800, Australia (email: nikola.zlatanov@monash.edu).} \thanks{H. Gačanin is with the Institute for Communication Technologies and Embedded Systems, RWTH Aachen University, 52074 Aachen, Germany (email: harisg@ice.rwth-aachen.de).} 
}

\markboth{}{Shell \MakeLowercase{\textit{et al.}}: Bare IEEEtran.cls for Journals} \maketitle

\begin{abstract}

In this paper, we propose a fairness-aware rate maximization scheme for a wireless powered communications network (WPCN) assisted by an intelligent reflecting surface (IRS). The proposed scheme combines user scheduling based on time division multiple access (TDMA) and (mechanical) angular displacement of the IRS. Each energy harvesting user (EHU) has dedicated time slots with optimized durations for energy harvesting and information transmission whereas, the phase matrix of the IRS is adjusted to focus its beam to a particular EHU. The proposed scheme exploits the fundamental dependence of the IRS channel path-loss on the angle between the IRS and the node's line-of-sight, which is often overlooked in the literature. Additionally, the network design can be optimized for large number of IRS unit cells, which is not the case with the computationally intensive state-of-the-art schemes. In fact, the EHUs can achieve significant rates at practical distances of several tens of meters to the base station (BS) only if the number of IRS unit cells is at least a few thousand. 

\end{abstract} 


\vspace{-5mm}

\section{Introduction}
Owing to the promise of perpetual and self-sustainable communication systems, wireless power transfer has attracted major research interest [\ref{ref-1}]-[\ref{ref-1b}]. However, the small form factors of the IoT devices impose small antenna apertures, which leads to tiny amounts of harvested RF energy and thus low achievable rates over practical communication ranges of the wireless powered communications networks (WPCNs). Fortunately, the recent concept of intelligent reflecting surfaces (IRS) can mitigate this issue by controlling and shaping the propagation environment between RF transmitters and energy harvesting (EH) users (EHUs). The IRS, an almost passive large metasurface with many sub-wavelength-sized elements acting as diffuse scatters, can relay and shape the signal from the RF transmitter (Tx) into a high-energy beam focused on the receiver (Rx). It therefore offers potential for great improvements in the spectrum and energy efficiency of future wireless networks [\ref{ref-3}]-[\ref{lit-new-2}], especially those employing short range line-of-sight (LoS) links such as the WPCNs. 

The existing literature on IRS-enabled WPCNs typically optimize the phase shifts of the IRS unit-cells (antenna elements), jointly with the precoding matrix at the base station (BS), the power allocation, and/or the time allocation [\ref{ref-6}]-[\ref{lit-new-3}]. In [\ref{ref-6}], the authors study an IRS-assisted downlink MIMO network with simultaneous wireless and information power transfer (SWIPT), where the system weighted sum-rate is iteratively maximized by alternating between first-order optimal points of the optimization sub-problems. The papers [\ref{ref-7}] and [\ref{ref-8}] optimize the BS beamformer and the IRS phase shifts to maximize the weighted sum power and the minimum power at the EHUs, respectively, by alternating between suboptimal algorithms obtained by semi-definite relaxation (SDR). Similarly, the paper [\ref{ref-9}] uses a successive alteration among the SDR, the Gaussian randomization, and the block coordinated descent method to arrive at the suboptimal solution for the IRS phase shift matrix and the time scheduling in a WPCN employing the TDMA protocol. Instead of TDMA, the paper [\ref{lit-new-1}] employs non orthogonal multiple access (NOMA), which introduces multiuser interference but results in coupled optimization variables and non-convex unit-modulus constraints. In [\ref{ref-10}], alternating optimization is proved unsuitable for optimizing a SWIPT-based MISO system with multiple IRSs, but instead utilize penalty-based (yet also suboptimal) algorithms. The paper [\ref{lit-new-3}] uses alternating optimization among power allocation, computation offloading and IRS phase matrix design to minimize energy consumption of a wireless powered mobile edge computing system, but results in a locally optimal solution. In addition to being suboptimal, the algorithms considered in the above mentioned literature are characterized by high computational complexity that grows exponentially with the number of IRS unit-cells. In fact, these algorithms are already computationally prohibitive for a few dozen unit-cells, while current IRS implementations consist of $10^4$ unit-cells [\ref{ref-14}].

Another critical issue is the applicability of the IRS channel models used commonly to optimize the IRS-assisted WPT systems. The channel from each user to the IRS is usually modelled as a conventional fading channel with random multi-path scattering. In fact, this channel can be modelled as a free-space deterministic channel [\ref{ref-11}]-[\ref{ref-14}], because an IRS is always deployed to maintain a LoS to each user and can form a narrow beam to that user. Additionally, [\ref{ref-6}]-[\ref{lit-new-3}] neglected that the deterministic path-loss depends on the specific angle between the IRS and the user's LoS but instead assume that users on the same distance from the IRS have identical average channel independent of the angle. This angular dependence can be used to introduce an additional degree of freedom into the design of the IRS-assisted system and thus improve its performance. 

To the best of authors’ knowledge, the mechanical tilt has not yet been applied in the context of the IRS. To get a better insight, in this paper we apply this functionality to a simple WPT system. We optimize the rate performance of a TDMA-based WPCN assisted by an IRS capable of angular displacement. For large number of unit-cells, the system exploits the capability of the IRS to focus very narrow beams towards a specific EHU. Instead of simultaneous WPT to all EHUs through a common EH phase, the EHUs are individually and successively powered by dedicated narrow energy beams ``steered" by the IRS.


\section{System Model} 

We consider a WPCN that consists of a BS, $K$ EHUs, and an IRS deployed on a location with a LoS to each node. All nodes are equipped with a single antenna and operate in half-duplex modes over the same frequency band by employing TDMA. The EHUs transmit information intended for the BS, whereas the BS operates as an information receiver and an energy beacon. The BS transmits at power $P_0$. Due to physical obstacles, the EHUs do not have direct LoSs to the BS, and thus the IRS facilitates the WPT and the information transmissions (IT) between the BS and the EHUs.\footnote{If the physical obstacle is absent, the inclusion of the direct link would result in a composite end-to-end channel between the BS and the $k$th EHU, which would suppress the characteristic impact of the mechanical tilt on the system performance. The stronger the gain of the direct component relative to the IRS-assisted channel, the smaller the impact of the mechanical tilt. } 

For maximum system performance, the horizontal orientation of the IRS can be mechanically tilted (c.f. Fig. 1). We assume that, relative to a reference direction ($x$-axis), the IRS is displaced by the angle $\Psi$, where $-\Psi_{N} \leq \Psi \leq \Psi_{P},$ such that $\Psi_{N}$ and $\Psi_{P}$ are the maximum angular displacements in negative (clockwise) and positive (counterclockwise) directions.  

\begin{figure}[t!]
 \begin{center}
  \includegraphics[trim = 55mm 195mm 4mm 20mm, scale=0.7]{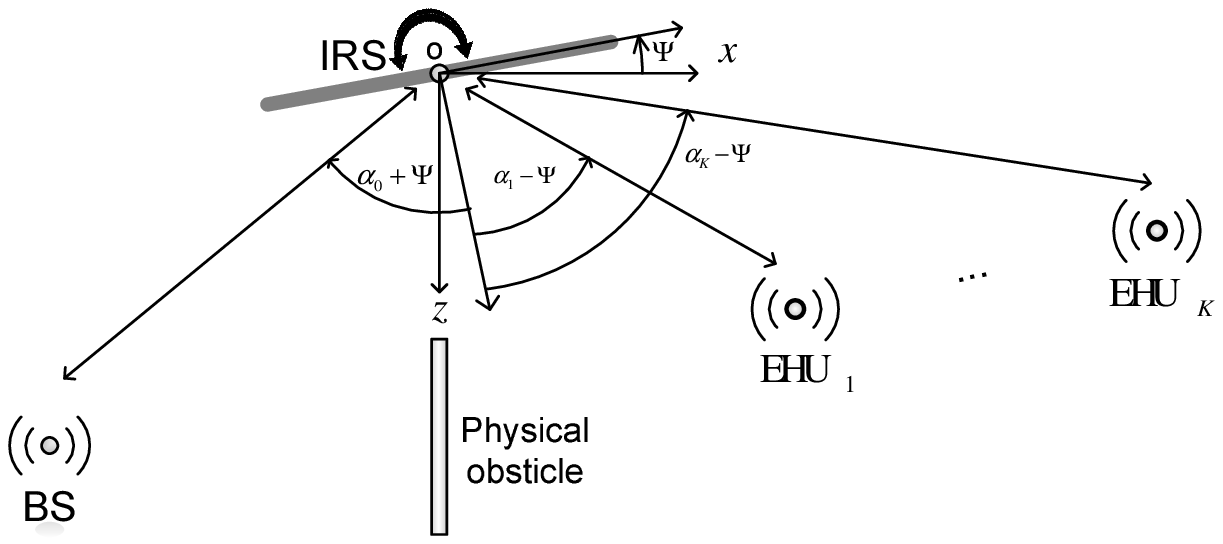}
 \end{center}
\caption{System model. } 
\label{fig1}
\end{figure}


\subsection{IRS and Channel Modeling} 
The IRS consists of $N$ unit-cells and has a total area $S$. It can be modelled as a planar antenna array, where each antenna element has an area of size $A$ satisfying $S = NA$ and $A \leq (\lambda/4)^2$ [\ref{ref-11}]-[\ref{ref-13}]. The reflection properties of the IRS are determined by the diagonal matrix $\mathbf{\Theta} = \text{diag} (e^{j \theta_1}, \cdots, e^{j \theta_N} )$, where $\theta_n \in [0, 2\pi )$ is the phase shift induced by the $n$th element of the IRS. These phase shifts are configurable and programmable via an IRS controller. 

Due to the existence of a LoS between the IRS and a given node, the channel gain between that node and the $n$th antenna element, $\Omega_n$, is deterministic and given by [\ref{ref-12}, Lemma $1$]. The respective channel phase, $\phi_n$, is also fixed and determined by the propagation delay over the distance $d_n$ between the node and the $n$th antenna element of the IRS, i.e., $\phi_n = 2 \pi d_{n}/\lambda$. Thus, the channel between the BS and the IRS is represented by the deterministic vector $\mathbf{h} = [h_1, \cdots, h_N]^T$, where $h_n = \sqrt{\Omega_{0n}} \, e^{-j \phi_{0n}}$ is the channel between the BS and the $n$th antenna element of the IRS. The channel between the IRS and the $k$th EHU ($1 \leq k \leq K$) is represented by the deterministic vector, $\mathbf{g}_k = [g_{k1}, \cdots, g_{kN}]^T$, where $g_{kn} = \sqrt{\Omega_{kn}} \, e^{-j \phi_{kn}}$ is the channel between the $k$th EHU and the $n$th antenna element of the IRS. Since the channels are deterministic and thus can be estimated arbitrarily well from pilot signals, the phases of all channels ($\phi_{0n}$ and $\phi_{kn}, \forall k, n$) are perfectly known by the IRS controller.  

\subsubsection*{Far-field approximation} In the remainder of the paper, we assume both the BS and the EHUs operate in the far-field region of the IRS. For unambiguous notation, all EHUs are assumed to be located in the first quadrant, while the BS is located in the second quadrant of the $x0z$ coordinate system (c.f. Fig. 1). The BS is located at polar coordinates $(d_0, -\alpha_0)$ with respect to (w.r.t) the pole $0$ and the $z$-axis, and so its angle w.r.t. the IRS boresight is $\alpha_0+\Psi$. The $k$th EHU is located at polar coordinates $(d_k, \alpha_k)$, and so its angle w.r.t the IRS boresight is $\alpha_k-\Psi$. Therefore, given $S \leq 9d_0^2$, the BS channel gains $\Omega_{0n}$ are approximated by (c.f. [\ref{ref-12}, Eq. (11) and Eq. (31)]) 
\begin{equation} \label{eq1}
    \Omega_{0n} \approx \frac{A \cos(\alpha_0+\Psi)}{4\pi d_0^2} \equiv \Omega_0 , \, \forall n,
\end{equation}
whereas channel gains of the $k$th EHU, $\Omega_{kn}$, are approximated by \footnote{Generally, the IRS channel can be modelled such that $\cos$-functions are of the form $\cos^q(\cdot)$, where the exponent $q$ is used to match the normalized power radiation pattern of an IRS unit-cell [\ref{ref-13}]. For example, the papers [\ref{ref-11}] and [\ref{ref-12}] assume $q = 1$, whereas the paper [\ref{ref-14}] assumes $q = 3$. We model the IRS channel according to [\ref{ref-12}]. The actual value of $q$ depends on the IRS (unit-cell) design, which is beyond the scope of this paper. Note, however, that increasing $q$ increases the {\it directivity} but decreases the {\it beamwidth} of the IRS beams towards the source (BS) and the destination ($k$th EHU). Thus, as $q$ increases, the effect of the mechanical tilt, $\Psi$, over the rate performance becomes more significant. On the other hand, if $q$ is low (e.g., $q \leq 0.5$), optimizing $\Psi$ has lesser impact on the system performance. Extremely, as $q \to 0$, the system is insensitive to the mechanical tilt. We actually use such a hypothetical IRS with $q=0$ in the numerical results section to establish a benchmark for estimating the impact of the mechanical tilt. }
\begin{equation} \label{eq2} 
     \Omega_{kn} \approx \frac{A \cos(\alpha_k-\Psi)}{4\pi d_k^2} \equiv  \Omega_k , \, \forall n.
\end{equation}
Without loss in generality, we may assume the EHUs are ordered by their increasing angular coordinates, i.e., $\alpha_1 < \alpha_2 < \cdots < \alpha_K.$ Thus, $\Psi_N = \pi/2 - \alpha_K$ and $\Psi_P = \pi/2 - \alpha_0$.


\subsection{Resource Allocation} 
The time during the communication session is divided into TDMA frames with unit durations. 
Each TDMA frame consists of $K$ time slots, where each slot is dedicated to a different EHU. We exploit the principal property of the IRS to focus its narrow beam to one (out of $K$) EHUs by using a dedicated phase shift design. For a large enough $N$, the IRS sharply focuses the beam on the $k$th EHU so that its radiation pattern rapidly decreases away from this EHU.\footnote{For analytical tractability of the proposed rate maximization, we neglect the RF energy harvested by those EHUs that are close to the ``targeted" EHU. The validity of this assumption grows stronger with increasing $N$. } Specifically, in the $k$th time slot ($1 \leq k \leq K$), the IRS focuses on the $k$th EHU by setting its reflection matrix, $\mathbf{\Theta}(k) = \text{diag} (e^{j \theta_1(k)}, \cdots, e^{j \theta_N(k)} )$, with the phase shifts: 
\begin{equation} \label{rev3_eq3} 
    \theta_n(k) = \phi_{0n} + \phi_{kn}, \qquad 1 \leq n \leq N.  
\end{equation}

The $k$th time slot is subdivided into an EH phase of duration $\nu_k$, and an IT phase of duration $\tau_k$. During the $k$th EH phase, the $k$th EHU is charged via the IRS by the downlink energy transmission from the BS. During the $k$th IT phase, the $k$th EHU spends the harvested energy to transmit information to the BS through an uplink transmission via the IRS. Note, in $k$th time slot, the IRS maintains the same reflection matrix, $\mathbf{\Theta}(k),$ for both the downlink and the uplink transmissions to/from the $k$th EHU. Assuming channel reciprocity, such phase shift design is consistent with [\ref{lit-new-1}, Proposition $1$]. 


Assuming a linear EH model, the energy harvested by the $k$th EHU in the $k$th EH phase is given by 
\begin{equation} \label{eq3}
    E_k = \eta_k P_0 \, \nu_k \, \lvert \mathbf{g}_k^T \mathbf{\Theta}(k) \mathbf{h} \rvert ^2 , 
\end{equation}
where $\eta_k$ denotes the (RF-to-DC) conversion efficiency of the $k$th EHU. In (\ref{eq3}), 
\begin{equation} \label{eq4}
    \lvert \mathbf{g}_k^T \mathbf{\Theta}(k) \mathbf{h} \rvert ^2  = \left| \sum_{n=1}^N \sqrt{\Omega_{0n} \Omega_{kn}} \right| ^2 = N^2 B_k(\Psi),
\end{equation} 
where 
\begin{equation} \label{eq5}
 B_k(\Psi) = \Omega_0 \Omega_k = \frac{A^2 \cos(\alpha_0+\Psi)\cos(\alpha_k-\Psi)}{(4\pi d_0 d_k)^2}.  
\end{equation} 
Note that (\ref{eq4}) corresponds to [\ref{ref-13}, Eq. (27)] and [\ref{ref-14}, Eq. (8)]. The achievable rate of the $k$th EHU is thus given by 
\begin{eqnarray} \label{eq6} 
    R_k &=& \tau_k \log \left(1 + \frac{E_k}{N_0 \tau_k} \lvert \mathbf{g}_k^T \mathbf{\Theta}(k) \mathbf{h} \rvert ^2 \right) \notag \\ &=& \tau_k \log \left(1 + N^4 B_k^2(\Psi) \frac{\eta_k P_0}{N_0} \frac{\nu_k}{\tau_k} \right), 
\end{eqnarray} 
where $N_0$ is the thermal noise power at the BS receiver. 

\vspace{-3mm}

\section{Common Rate Maximization} 

We aim at maximizing the minimal rate of all EHUs, $\max \{R_k\}_{k=1}^K$, by the optimal adjustment of the horizontal displacement angle of the IRS and of the durations of the EH and IT phases of all EHUs. This criterion guarantees fair resource sharing and a minimal common rate to all EHUs, $R_0$, thus successfully tackling the network's near-far problem. The optimization problem is specifically stated as 
\begin{equation} \label{op}
\underset{R_0, \Psi, \, \nu_{k}, \tau_k, \forall k} {\text{Maximize}} \ R_0  \notag 
\end{equation}
\vspace{-0mm}
\hspace{20mm}\text{subject to: } 
\begin{eqnarray}
\begin{array}{ll}
     C1: & R_k \geq R_0, \forall k \\
     C2: & \sum_{k = 1}^K \left( \nu_k + \tau_k \right) =1\\
     C3: & -\Psi_{N}\leq \Psi\leq \Psi_{P}\\
     C4: & \nu_k \geq 0, \, \tau_k \geq 0, \, \forall k. 
\end{array}
\end{eqnarray}
The problem in (\ref{op}) is not convex, but can be solved by splitting it into two subproblems, one of which is convex. Specifically, for a given $\Psi_0$, (\ref{op}) is convex with respect to the optimization variables $R_0, \nu_{k}, \tau_k, \forall k$, which can be determined analytically in closed form according to the following theorem. 

\begin{theorem}
Let us denote $C_k = \eta_k P_0 N^4 B_k^2(\Psi) / N_0$. The common rate of all EHUs in the considered system is determined by 
\begin{equation} \label{op_value}
R_0^* = \left[\sum_{k=1}^K\frac{1-1/C_k}{W\left(\frac{C_k-1}{e}\right)}   \right]^{-1} \, ,
\end{equation}
where $W(\cdot)$ is the Lambert $W$ function. The optimal duration of the IT phase of the $k$th EHU is determined by 
\begin{equation} \label{op_tau}
    \tau_k^* = R_0^* \, \left[1+W\left(\frac{C_k-1}{e}\right)\right]^{-1}, 
\end{equation}
whereas the optimal duration of the EH phase of the $k$th EHU is determined by 
\begin{equation} \label{nu_opt}
    \nu_k^* = \frac{\tau_k^*}{C_k}\left[\frac{C_k-1}{W\left(\frac{C_k-1}{e}\right)}-1\right].
\end{equation}
\end{theorem}
\begin{IEEEproof}
Please refer to Appendix A.
\end{IEEEproof} 
Given Theorem $1$, the optimal value of the IRS angular displacement is determined by 
\begin{equation} \label{op_out}
\Psi^* = \underset{-\Psi_{N}\leq \Psi\leq \Psi_{P}} {\text{argmax}} R_0^* \, ,  
\end{equation}
where $R_0^*$ is given by (\ref{op_value}). In practice, (\ref{op_out}) can be tackled numerically by some of the well known algorithms for single variable optimization, such as, the bisection method, the Newton's method, or the secant method.

\section{Numerical results}

In this section, we illustrate the performance of the proposed system design and resource allocation for the IRS-assisted WPCN. We specifically study the sum rate, $R_{sum} = \sum_{k=1}^K R_k$, for two system settings, denoted by "Optimal $\Psi$" and "$\Psi=0$", respectively. The setting "Optimal $\Psi$" refers to the WPCN resource allocation per Theorem $1$ at the optimal angular displacement of the IRS, determined by (\ref{op_out}). The setting "$\Psi = 0$" refers to the WPCN resource allocation per Theorem $1$ at zero angular displacement of the IRS. 

Each unit-cell of the IRS has an area $A = (\lambda/4)^2$ with $\lambda = 0.1$m (i.e., carrier frequency of $3$GHz). Unless stated otherwise, the IRS consists of $N = 10^4$ unit-cells. The BS transmit power is set to $P_0 = 4$W, and the thermal noise power is set to $N_0 = 10^{-13}$W. The EHUs are uniformly distributed along an arc of radius $20$m in the first quadrant of the $x0z$ coordinate system within a range of polar angles $[\pi/40, \, 19\pi/40]$. Additionally, $\eta_k = 0.9, \forall k$, for all the EHUs. 

\subsubsection*{Benchmark IRS} We consider a {\it benchmark IRS} whose radiation pattern of its unit-cells is insensitive to its angle with the node at which the IRS focuses its beam. Such IRS behaves similarly to a semi-directional antenna, such as, patch/panel or sector antenna, with (hypothetical) beamwidth of 180 degrees. To satisfy the conservation of power (c.f. [\ref{ref-13}, Eq. (16)], [\ref{ref-14}, Eq. (2)]), the gain of an IRS unit-cell is halved compared to the gain of the (practical) IRS considered in Section II. Thus, for the benchmark IRS, the gains of the channels from its $n$th unit-cell to the BS and to the $k$th EHU are modelled as $\hat \Omega_{0n} = A/(8 \pi d_0^2)$ and $\hat \Omega_{kn} = A/(8 \pi d_k^2)$, respectively. When employing the benchmark IRS, the EHUs' common rate is again maximized by the resource sharing proposed by Theorem $1$, which is however insensitive to $\Psi.$  

\begin{figure}[t!]
 \begin{center}
  \includegraphics[trim = 5mm 0mm 0mm 0mm, scale=0.47]{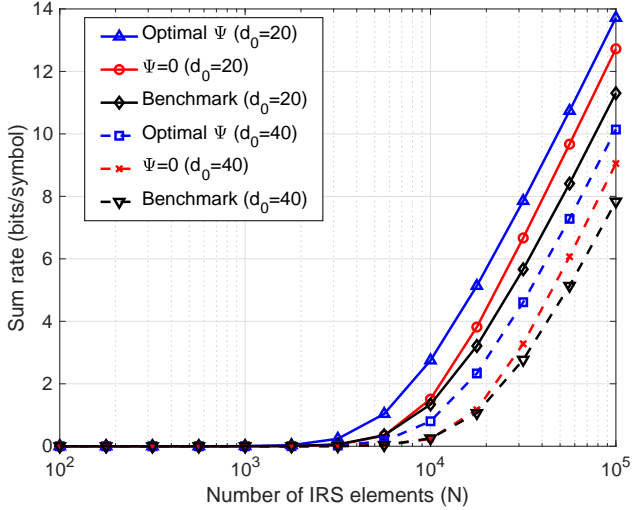}
 \end{center}
 \vspace{-3mm} 
\caption{WPCN sum rate vs. $N$ for $K = 10$. When the nodes are $20$m or more away from the IRS, WPCN attains meaningful sum rate if the number of IRS unit-cells is in the thousands. } 
\label{fig2b}
\end{figure}

Assuming $K = 10$ EHUs and $\alpha_0 = \pi/12$, Fig. \ref{fig2b} depicts $R_{sum}$ vs. $N$ for $d_0 = 20$m and $d_0 = 40$m. The considered system attains meaningful rates when the IRS is comprised of at least several thousand unit-cells. In this region, mechanical tilt, if set properly, results in an evident rate improvement, which is nearly independent of $N$. On the other hand, an IRS with $N \leq 10^3$ is useful only if all EHUs are located near the IRS. For the given settings, the system with the benchmark IRS performs worse then the system with the practical IRS. 

\begin{figure}[t!]
 \begin{center}
  \includegraphics[trim = 5mm 0mm 0mm 0mm, scale=0.47]{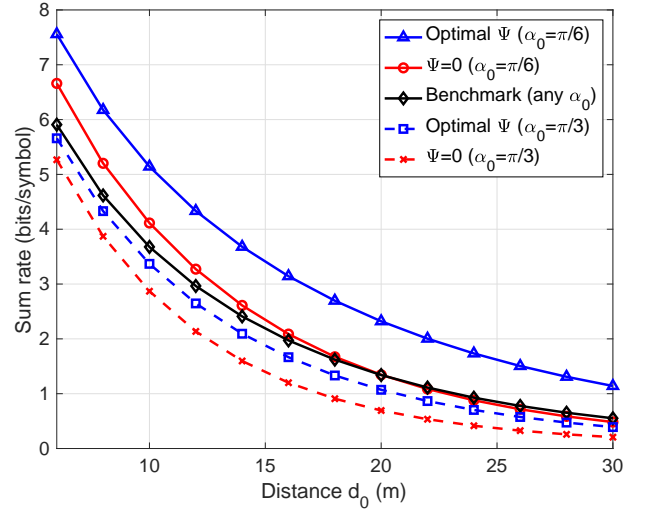}
 \end{center} 
 \vspace{-3mm} 
\caption{WPCN sum rate vs. BS-IRS distance for $N=10^4$ and $K=10$. The rate improvement with optimal mechanical tilt is almost constant relative to $d_0$. } 
\label{fig3a}
\end{figure}

The sum rate generally strongly depends on both $d_0$ and $\alpha_0$. Fig. \ref{fig3a} depicts $R_{sum}$ vs. $d_0$ for $\alpha_0 = \pi/6$ and $\alpha_0 = \pi/3$. The rate improvement obtained by the mechanical tilt of the IRS is almost constant in the considered range of $d_0$. Here, the benchmark IRS performs worse than the practical IRS with $\alpha_0 = \pi/6$, but better than the practical IRS with $\alpha_0 = \pi/3$.

\begin{figure}[t!]
 \begin{center}
  \includegraphics[trim = 5mm 0mm 0mm 0mm, scale=0.47]{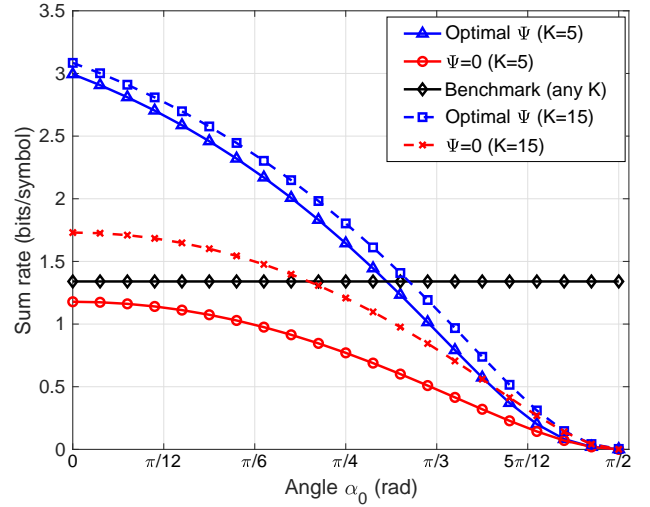}
 \end{center} 
 \vspace{-3mm}
\caption{WPCN sum rate vs. $\alpha_0$ for $d_0 = 20m$. The WPCN rate under optimal mechanical tilt is almost independent of the number of EHUs. } 
\label{fig4b}
\end{figure}

Fig. \ref{fig4b} investigates the influence of $\alpha_0$. The sum rate decreases with increasing $\alpha_0$ due to the decreasing effective area of the IRS with respect to the BS (c.f. (1)). The benefit from the mechanical tilt is highest when $\alpha_0 \approx 0$. In this case, the ``Optimal $\Psi$" setting decreases $\Omega_0$ of the BS-IRS channel, but increases $\Omega_k$ of almost all EHU-IRS channels. This trade-off overall yields sum rate improvement because even the weakest EHU-IRS channel is improved by the  mechanical tilt. Fig. \ref{fig4b} also shows that the sum rate of the ``Optimal $\Psi$" setting is marginally affected by the number of EHUs, while the impact of $N$ is more significant when $\Psi=0$. Actually, the mechanical tilt with $\Psi = \Psi^*$ smooths out the differences between the gains of different EHU-IRS channels. In this case, $\tau_k^*$ (i.e., $\nu_k^*$) of all EHUs attain similar values. Conversely, when $\Psi=0$, the EHUs with $\alpha_k \approx 0$ will have much better channels compared to the EHUs with $\alpha_k \approx \pi/2$, and so the values of $\tau_k^*$ (i.e., $\nu_k^*$) will have greater variance. In this case, the sum rate is increased by adding more EHUs. If the benchmark IRS were employed, the sum rate would be fixed regardless of $\alpha_0.$ Up to some threshold $\alpha_0$, its sum rate would be below that of the practical IRS, but would exceed it beyond this threshold. In this case, the system with a BS placed normally to the IRS boresight would perform identically to a system with a BS placed along the IRS boresight. Such behaviour of the benchmark IRS clearly emphasizes the importance of the proper modeling of the IRS channel. Actually, the performance of the system employing a practical IRS is significantly affected by the angle between the IRS and the nodes, which is often neglected in the literature.

\section{Conclusion} 

In this paper, we proposed a practical resource allocation scheme that maximizes the minimum common rate of a WPCN assisted by an IRS with mechanically adjustable horizontal orientation. The common rate maximization guarantees a fair bandwidth sharing among the EHUs. The phase matrix of the IRS is successively adjusted from one time slot to the next to direct its beam toward each EHU. Unlike the schemes available in the literature, the system under consideration can be optimized for the IRS with a large number of unit-cells, which is consistent with the state-of-the-art IRS implementations.


\appendices
\section{Proof of Theorem 1}
For given $\Psi_0$, the problem (\ref{op}) is convex, because the objective function is affine and the constraints $C1$ and $C2$ are jointly convex with respect to $R_0$, $\tau_k$, and $\nu_k$, $\forall k$. Thus, we can apply the Lagrangian dual method to solve the original problem for a fixed $\Psi_0$. The Lagrangian of (\ref{op}) is given by 
\begin{eqnarray} \label{eq_lagr}
\notag
\mathcal{L} \hspace{-1mm}&=&\hspace{-1mm} R_0 + \sum_{k=1}^K\lambda_k\left(R_0-\tau_k\log\left(1+C_k\frac{\nu_k}{\tau_k}\right)\right)\\
&-&\lambda_0\left(\sum_{k=1}^K\tau_k+\nu_k-1\right)\\ \notag
\end{eqnarray}
where $\lambda_k$ and $\lambda_0$ are the non-negative Lagrangian multipliers associated with the constraints $C_1$ and $C_2$, respectively. After differentiating $\mathcal{L}$ with respect to $R_0$, $\tau_k$ and $\nu_k$ and setting the derivatives to zero, we obtain: 
\vspace{0mm}
\begin{equation} \label{der_R0}
1-\sum_{k=1}^K\lambda_k = 0 \, ,
\vspace{0mm}
\end{equation}
\begin{equation} \label{der_tauk} 
\lambda_k\left(\log\left(1+C_k\frac{\nu_k}{\tau_k}\right)-\frac{C_k\frac{\nu_k}{\tau_k}}{1+C_k\frac{\nu_k}{\tau_k}}\right)-\lambda_0 = 0, \,\forall k, 
\vspace{0mm}
\end{equation}
\begin{equation} \label{der_nuk}
 \frac{C_k\lambda_k}{1+C_k \nu_k/\tau_k}-\lambda_0 = 0, \, \forall k.  
\vspace{0mm}
\end{equation}
Next, we introduce an auxiliary variable 
\begin{equation} \label{xk_prv}
x_k = C_k\frac{\nu_k}{\tau_k}, 
\end{equation}
and so (\ref{der_nuk}) is transformed as  
\begin{equation} \label{eq_xk}
    x_k = C_k\frac{\lambda_k}{\lambda_0} - 1. 
\end{equation}
Introducing (\ref{eq_xk}) in (\ref{der_tauk}) leads to the transcendental equation, 
\begin{equation} \label{eq_lambdi}
    \log\left(C_k\frac{\lambda_k}{\lambda_0}\right) + \left(\frac{1}{C_k}-1\right)\frac{\lambda_0}{\lambda_k}-1=0 , 
\end{equation}
which can be solved in closed form by applying the properties of the Lambert-$W$ function [\ref{ref-1b}], as 
\begin{equation}
    \label{eq_lambdi_resh}
    \frac{\lambda_k}{\lambda_0} = \left(1-\frac{1}{C_k} \right)\left[W\left(\frac{C_k-1}{e} \right)\right]^{-1} . 
\end{equation}
Due to the convexity of the considered problem, all the constraints $C_1$ are satisfied with strict equality, yielding 

\begin{equation} \label{eq_tau_k}
    \tau_k = \frac{R_0}{\log(1+x_k)}. 
\end{equation}
Applying (\ref{xk_prv}), (\ref{eq_xk}), (\ref{eq_lambdi_resh}) and (\ref{eq_tau_k}), the left hand side of $C_2$ is transformed as 
\begin{eqnarray} \label{eq_R0}
 \sum_{k=1}^K\tau_k+\nu_k 
 \overset{(a)}{=} R_0\sum_{k=1}^K\frac{\left(1-\frac{1}{C_k}\right)\left(1+\frac{1}{W\left(\frac{C_k-1}{e} \right)}\right)}{\log\left((C_k-1)/W\left(\frac{C_k-1}{e} \right)\right)} \nonumber \\
 \overset{(b)}{=}R_0\sum_{k=1}^K\frac{\left(1-\frac{1}{C_k}\right)\left(1+\frac{1}{W\left(\frac{C_k-1}{e} \right)}\right)}{1+W\left(\frac{C_k-1}{e} \right)},  
\end{eqnarray}
where $(a)$ results from applying (\ref{eq_xk}) and (\ref{eq_lambdi_resh}), and $(b)$ results from $W$-function definition, $x = W(x) \, e^{W(x)}$. Thus, setting (\ref{eq_R0}) to unity yields (\ref{op_value}); Inserting (\ref{eq_xk}) and (\ref{eq_lambdi_resh}) into (\ref{eq_tau_k}) yields (\ref{op_tau}); Inserting (\ref{op_tau}), (\ref{eq_xk}) and (\ref{eq_lambdi_resh}) into (\ref{xk_prv}) yields (\ref{nu_opt}).




\begin{thebibliography}{99}


\bibitem{ref-1} I. Krikidis, S. Timotheou, S. Nikolaou, G. Zheng, D. W. K. Ng and R. Schober, ``Simultaneous wireless information and power transfer in modern communication systems," \textit{IEEE Commun. Mag.}, vol. 52, no. 11, pp. 104-110, Nov. 2014 \label{ref-1} 

\bibitem{ref-1a} H. Ju and R. Zhang, ``Throughput maximization in wireless powered communication networks," \textit{IEEE Trans. Wireless Commun.}, vol. 13, no. 1, pp. 418-428, Jan. 2014  \label{ref-1a} 

\bibitem{ref-1b} S. Pejoski, Z. Hadzi-Velkov and R. Schober, ``Optimal power and time sllocation for WPCNs with piece-wise linear EH model," \textit{IEEE Wireless Commun. Let.}, vol. 7, no. 3, pp. 364-367, Jun. 2018
\label{ref-1b} 

\bibitem{ref-3} E. Basar, M. Di Renzo, J. De Rosny, M. Debbah, M. Alouini and R. Zhang, ``Wireless communications through reconfigurable intelligent surfaces," \textit{IEEE Access}, vol. 7, pp. 116753-116773, 2019  \label{ref-3} 

\bibitem{lit-new-2} C. Pan et al., ``Multicell MIMO communications relying on intelligent reflecting surfaces," \textit{IEEE Trans. Wireless Commun.}, vol. 19, no. 8, pp. 5218-5233, Aug. 2020 \label{lit-new-2} 

\bibitem{ref-6} C. Pan et al., ``Intelligent reflecting surface aided MIMO broadcasting for simultaneous wireless information and power transfer," \textit{IEEE J. Sel. Areas Commun.}, vol. 38, no. 8, pp. 1719-1734, Aug. 2020 \label{ref-6}  

\bibitem{ref-7} Q. Wu and R. Zhang, ``Weighted sum power maximization for intelligent reflecting surface aided SWIPT," \textit{IEEE Wireless Commun. Let.}, vol. 9, no. 5, pp. 586-590, May 2020 \label{ref-7} 

\bibitem{ref-8} Y. Tang, G. Ma, H. Xie, J. Xu and X. Han, ``Joint transmit and reflective beamforming design for IRS-assisted multiuser MISO SWIPT systems," \textit{IEEE ICC 2020}, Dublin, Ireland, 2020, pp. 1-6 \label{ref-8}  

\bibitem{ref-9} B. Lyu, D. T. Hoang, S. Gong and Z. Yang, ``Intelligent reflecting surface assisted wireless powered communication networks," \textit{Proc. WCNC Workshops 2020}, Seoul, Korea, 2020, pp. 1-6 \label{ref-9} 

\bibitem{lit-new-1} Q. Wu, X. Zhou and R. Schober, ``IRS-assisted wireless powered NOMA: Do we really need different phase shifts in DL and UL?," \textit{IEEE Wireless Commun. Let.}, early access, doi: 10.1109/LWC.2021.3072502 \label{lit-new-1}

\bibitem{ref-10} Q. Wu and R. Zhang, ``Joint active and passive beamforming optimization for intelligent reflecting surface assisted SWIPT under QoS constraints," \textit{IEEE J. Sel. Areas Commun.}, vol. 38, no. 8, pp. 1735-1748, Aug. 2020 \label{ref-10}  

\bibitem{lit-new-3} T. Bai at al., ``Resource allocation for intelligent reflecting surface aided wireless powered mobile edge computing in OFDM systems," \textit{IEEE Trans. Wireless Commun.}, early access, doi: 10.1109/ TWC.2021.3067709  \label{lit-new-3} 


\bibitem{ref-11} Ö. Özdogan, E. Björnson and E. G. Larsson, ``Intelligent reflecting surfaces: physics, propagation, and pathloss modeling," \textit{IEEE Wireless Commun. Let.}, vol. 9, no. 5, pp. 581-585, May 2020 \label{ref-11}  

\bibitem{ref-12} E. Björnson and L. Sanguinetti, ``Power scaling laws and near-field behaviors of massive MIMO and intelligent reflecting surfaces," \textit{IEEE Open J. Commun. Soc.}, vol. 1, pp. 1306-1324, 2020 \label{ref-12}  

\bibitem{ref-13} S. W. Ellingson, ``Path loss in reconfigurable intelligent surface-enabled channels," \textit{arXiv}, Dec. 2019. [Online]. Available: https://arxiv.org/abs/1912.06759 \label{ref-13}  

\bibitem{ref-14} W. Tang et al., ``Wireless communications with reconfigurable intelligent surface: path loss modeling and experimental measurement," \textit{IEEE Trans. Wireless Commun.}, vol. 20, no. 1, pp. 421-439, Jan. 2021 \label{ref-14}  



\end{thebibliography}
\end{document}